\documentclass[letter]{aa} 
\usepackage{graphicx}
\usepackage{txfonts}
\usepackage{natbib}
\bibpunct{(}{)}{,}{a}{}{,}
\def\Mspy   {\ifmmode {M_{\odot} {\rm yr}^{-1}} \else $M_{\odot}$~yr$^{-1}$\fi}
\def\Mdot   {\ifmmode {\dot M} \else $\dot M$\fi}
\def\as     {\ifmmode {\rlap.}$\,$''$\,$\! \else ${\rlap.}$\,$''$\,$\!$\fi}
\def\decsec  {\ifmmode {\rlap.}$\,$^{s}$\,$\! \else ${\rlap.}$\,$^{s}$\,$\!$\fi}\def\decs  {\ifmmode {\rlap.}$\,$^{s}$\,$\! \else ${\rlap.}$\,$^{s}$\,$\!$\fi}

\newcommand{\kms}{\mbox{km~s$^{-1}$}}

\newbox\grsign      \setbox\grsign=\hbox{$>$} 
\newdimen\grdimen   \grdimen=\ht\grsign
\newbox\simgreatbox \setbox\simgreatbox=\hbox{\raise.5ex\hbox{$>$}\llap
                        {\lower.5ex\hbox{$\sim$}}}\ht1=\grdimen\dp1=0pt
\newbox\simlessbox  \setbox\simlessbox =\hbox{\raise.5ex\hbox{$<$}\llap
                        {\lower.5ex\hbox{$\sim$}}}\ht2=\grdimen\dp2=0pt

\newcommand{\uchii}{UCHII}
\newcommand{\uchiir}{UCHIIR}
\begin{document}
\title{SOFIA observations of far-infrared hydroxyl emission toward classical ultracompact HII/OH maser regions}
 \subtitle{}

 \author{T. Csengeri
 \inst{1}
 \and
 K. M. Menten
 \inst{1}
 \and
F. Wyrowski
\inst{1}
\and
M. A. Requena-Torres
\inst{1}
\and
 R. G{\" u}sten
 \inst{1}
\and
H. Wiesemeyer
        \inst{1}
\and
H.-W. H\"ubers
       \inst{2,3}    
\and
P. Hartogh
\inst{4}
\and
K. Jacobs
\inst{5} 
             }


 \institute{   
Max-Planck-Institut f{\"u}r Radioastronomie, Auf dem H{\"u}gel 69,
D-53121 Bonn, Germany 
\email{[ctimea;kmenten]@mpifr.de}
\and
Deutsches Zentrum f\"ur Luft- und Raumfahrt, Institut f¬ur Planetenforschung, Rutherfordstra\ss e 2, 12489 Berlin, Germany
\and
Institut f\"ur Optik und Atomare Physik, Technische Universit\"at Berlin, 
Hardenbergstra\ss e 36, 10623 Berlin, Germany
\and 
Max-Planck-Institut f\"ur Sonnensystemforschung, 
Max-Planck-Stra\ss e 2, 37191 Katlenburg-Lindau, Germany
\and
I. Physikalisches Institut, Universitt zu K\"oln, Z\"ulpicher Str. 77, 50937 K\"oln, Germany
}

 \date{Received ... ; accepted}
\titlerunning{SOFIA observations of OH toward classical {\uchii} regions}
\abstract
{The hydroxyl radical (OH) is found in various environments within the interstellar medium (ISM) of the Milky Way and external galaxies, mostly either in diffuse interstellar clouds or in the warm, dense environments of newly formed low-mass and high-mass stars, i.e, in the dense shells of compact and ultracompact HII regions (UCHIIRs).
Until today, most studies of interstellar OH involved the molecule's radio wavelength hyperfine structure (hfs) transitions. 
These lines are generally not in local thermodynamical equilibrium (LTE) and either masing or over-cooling complicates their interpretation. In the past, observations of transitions between different rotational levels of OH, which are at far-infrared wavelengths, 
have suffered from limited spectral and angular resolution. Since these lines have critical densities many orders of magnitude higher than the radio wavelength ground state hfs lines and are emitted from levels with more than 100 K above the ground state, when observed in \textit{emission}, they probe very dense and warm material.}
{We aim to probe the warm and dense molecular material surrounding the UCHIIR/OH maser sources W3(OH), G10.62$-$0.39 and NGC~7538~IRS1 by studying  the $^2\Pi_{{1/2}}, J  = {3/2} - {1/2}$ rotational transition of OH in emission and, toward the last source also the molecule's $^2\Pi_{3/2}, J = 5/2 - 3/2$ ground-state transition in absorption. 
}
{We used the Stratospheric Observatory for Infrared Astronomy (SOFIA) to observe these OH lines, which are near 1.84 THz  ($163~\mu$m) and 2.51 THz  ($119.3~\mu$m), with high angular 
($\sim$16''/11'') and
spectral resolution ({better than 1 km~s}$^{-1}$).} 
{We clearly detect the OH lines, some of which are blended with each other. Employing non-LTE radiative transfer calculations we predict line intensities using models of a low OH abundance envelope versus a compact, high-abundance source corresponding to the origin of the radio OH lines. From the observed velocities and line-widths we can place constraints on the origin of the emission, and with detailed modeling we show for instance that the OH emission of W3(OH) comes from the {\uchiir} and not from the envelope of the nearby hot-core.
}
{
The far-IR lines of the OH molecule provide important information on the density and temperature structure of {\uchiir}s. Taking a low-abundance envelope component is not sufficient to reproduce the spectra for W3(OH) and G10.62$-$0.39, a compact, high OH column density source -- corresponding to the OH radio emitting sources is definitively required.
}


 \maketitle
%

\section{Introduction}
The hydroxyl radical (OH) was the first interstellar molecule detected at radio wavelengths. 
After various searches, \citet{Weinreb1963} found absorption in the two strongest ($F = 2 - 2$ and $2-1$, near 1667 and 1665 MHz, respectively) of the hyperfine structure (hfs) transitions within the molecule's $^2\Pi_{3/2}, J = 3/2$ ground state toward the supernova remnant Cas A. Subsequent studies found absorption toward all four of the ``18 cm lines''  first toward the Galactic center \citep{Gardner1964} and, later, many galactic lines of sight 
\cite[e.g.,][]{Goss1968}. Soon, also \textit{emission} near 1665 MHz was discovered toward prominent HII regions, i.e., sites of recent star formation \citep{Weaver1965}. This and the detection of weaker emission at the frequencies of the $F = 2-2$ and $1-2$  OH hfs lines (at 1667 and 1720 MHz) allowed \citet{Weinreb1965} to identify of OH as the carrier of this peculiar emission. It was readily identified as due to maser action by  the latter authors because of the observed peculiar profiles with multiple, narrow velocity components of polarized emission with highly anomalous relative hfs component intensity ratios, with the 1665 MHz $F = 2-1$ line usually being the strongest \citep{Weinreb1965, BarrettRogers1966}. The maser nature was clinched by interferometry, which showed that the OH emission near the prominent HII region W3 had brightness temperatures exceeding $10^9$~K and arose from compact ($<0{\rlap.}''05$) spots {spread over a few arc sec (thousands of AU) and significantly offset (by $14'$ or $\approx 8$ pc)} from the radio continuum emission maximum of the HII region \citep{Rogers1966,Moran1967,Davies1967}. Most interestingly, {the maser position} turned out to be consistent with that of {a newly detected radio source much smaller than W3:} the archetypical example of what \citet{Mezger1967},  based on their 2 cm radio continuum observations, termed \textit{``a new class of compact H II regions associated with OH emission sources''.} 
 This source, W3(OH), is characterized by its youth \citep[2300~yr, ][]{Kawamura1998}, compactness ($\sim 0.02$ pc) and high electron densities \citep[$\sim 2~10^5$~cm$^{-3}$,][]{DreherWelch1981} and is still surrounded by an expanding dusty, molecular envelope \citep[e.g.,\,][]{Wyrowski1999}. It is the archetypical ultracompact HII region (UCHIIR) and is discussed in this \textit{Letter} along with the {\uchii}Rs G10.62$-$0.39 and NGC 7538 IRS1. 

In addition to the lines from the ground state, transitions between hfs levels of various rotationally excited 
states with rotational quantum number $J$ from  $5/2$ to $9/2$ from the lower energy $^2\Pi_{3/2}$ ladder
and from $1/2$ to $5/2$ in the $^2\Pi_{1/2}$ ladder have been 
detected at radio frequencies between 4 and 23 GHz \citep[see][for a summary of the observations and 
modeling]{CW91}. These lines 
probe 
levels with energies up to 511~K above ground, whose populations are determined by an interplay of 
strong far-infrared (FIR) 
radiation from warm dust and collisions in dense gas since the FIR rotational lines that (de)populate them, at frequencies $>1.8$~THz, have high critical 
densities. Radiative transfer calculations show that many of them are not in local thermodynamic equilibrium (LTE), which 
gives rise to maser emission in some lines and enhanced absorption, caused by ``overcooling'',  in others, see
 \citet[][]{CW91}. As these authors show, just the fact that some lines do show maser emission and 
 others do not places interesting constraints on kinetic temperature and density.


The modeling of the various cm OH hfs lines delivers a consistent picture of the OH in warm, dense gas. Nevertheless, for a full characterization, observations of the \textit{rotational} OH transitions connecting hfs levels from different rotationally excited states are desirable.  The frequencies of these transitions, within the $^2\Pi_{3/2}$ and $1/2$ ladders, and also between them, 
all lie in the  supra-THz range. 
Using the Kuiper Airborne Observatory (KAO), \citet{Storey1981} made the first detection of the $^2\Pi_{3/2}, J = 5/2 - 3/2$ ground-state transitions near 2.51 THz  ($119.3~\mu$m) in absorption toward Sgr B2 and in emission from the shocked gas in the Kleinmann-Low (KL) nebula in the Orion molecular could 1. Detections of other OH lines followed, including those of the other lines discussed in the present \textit{Letter}, the $^2\Pi_{1/2}, J=3/2 - 1/2$ transitions near 1.83 THz (164~$\mu$m), also in Orion \citep{Viscuso1985}. 
The Infrared Observatory (ISO) brought a significant improvement in sensitivity, although neither in angular nor in spectral resolution.  Again, studies concentrated on the exceptionally bright and extended sources Sgr B2 and Orion-KL. Multi-transition data afforded by ISO's long wavelength spectrometer allowed detailed excitation studies of both sources in several hydroxyl isotopologes ($^{16, 17, 18} $OH) (\citealp{Goico2002}; \citealp{Goico2006}; \citealp{Polehampton2003}).

Herschel and SOFIA 
have brought vast improvements over both the KAO and ISO in terms of sensitivity and angular and velocity resolution. 
Both are flying heterodyne spectrometers affording bandwidths of several hundred km~s$^{-1}$ even at THz frequencies and velocity resolutions better than 1 km~s$^{-1}$, namely the Heterodyne Instrument for the Far Infrared \citep[HIFI,][]{deGraauw2010} and the German Receiver for Astronomy at Terahertz frequencies (GREAT\footnote{German Receiver 
for Astronomy at Terahertz frequencies. GREAT is a development by the MPI f\"ur 
Radioastronomie and the KOSMA/Universit\"at zu K\"oln, in cooperation 
with the MPI f\"ur Sonnensystemforschung and the DLR Institut 
f\"ur Planetenforschung.}, \citealp{Heyminck}).
The increased sensitivity allows studies of more compact regions, namely the above mentioned dense molecular envelopes associated with UCHIIRs and OH masers. Recently, \citet{Wampfler2011} presented HIFI observations  of the  $^2\Pi_{1/2}, J=3/2 - 1/2$ triplet of OH hfs transitions at 1837.8 GHz (163.1 $\mu$m) toward the high-mass star-forming region W3 IRS 5, for the first time resolving the hfs structure. A multi-rotational OH line study of the Orion Bar photon-dominated region with the PACS instrument onboard Herschel was presented by \citet{Goico2011}. Here we report in Sec. \ref{obs} our SOFIA observations of emission in this and the %
other $\Lambda$-doublet component at 1834.7 GHz (163.4~$\mu$m) toward the well-known UCHIIRs/OH maser sources W3(OH), G10.62$-$0.39 and NGC7538. The latter two have similar warm, dense molecular environments, and also show radio OH maser emission \citep{Argon2000} and absorption \citep{Walmsley1986}. Toward the latter source we also observed absorption from the $^2\Pi_{3/2}, J = 5/2 - 3/2$ ground-state transitions near 2.51 THz  ($119.3~\mu$m). 

\section{\label{obs}Observation and data reduction}
The OH $^2\Pi_{1/2}, J=3/2 - 1/2$ rotational lines were 
observed with the PI-instrument GREAT onboard SOFIA. 
The observations were carried out during flights no. SS02--OCF4/03 and
BS02--01/06/07. The signal was detected at 1837.8~GHz (163.1~$\mu$m)
using the L2 band of GREAT, while the pair of transitions from the other  $\Lambda$-doublet 
at 1834.7 GHz (163.4~$\mu$m) lies in the image sideband. The beam is 
$15{\rlap.{}''}5$ at 1837.8~GHz. 
The AFFT backend was used with 1.5 GHz bandwidth and 212 kHz spectral resolution.
The average system temperature is 4700--5300 K. Toward NGC~7538~IRS1  the $^2\Pi_{3/2}, J=5/2^{-} - 3/2^{+}$ rotational line at 2514~GHz was observed using the
M-band receiver. See 
\citet{Wiesemeyer2012} for details of these observations.
The pointing was established with the optical guide cameras (with an accuracy of 5{\arcsec}). 

The data were calibrated using standard
procedures based on the KOSMA/GREAT calibrator \citep{XinG}, for the data analysis  
the GILDAS\footnote{See http://www.iram.fr/IRAMFR/GILDAS/} software
was used. The forward efficiency was set to 95\% 
and the data were converted to a Rayleigh-Jeans equivalent 
T$_{\rm mb,RJ}$ scale 
using a beam efficiency of 51\% for the L2 and 58\% for the M band, respectively. 

The spectra were first summed up to define windows 
excluding the line signal for the baseline fitting. Then a 
third order polynomial baseline was fitted and subtracted from each scan 
individually. The $^2\Pi_{1/2}, J=3/2 - 1/2$ spectra were averaged with noise weighting and 
are smoothed to $\sim$1.2 \kms\,. 
We reach on average $\sim$0.3--0.4~K noise level per channel. The $^2\Pi_{3/2}, J=5/2^{-} - 3/2^{+}$
line was reduced in a similar fashion. Here, the spectrum is smoothed to 0.9~\kms\
velocity resolution and has a 0.4~K noise level.

\begin{table*}
\caption{Summary of the derived line parameters of the $^2\Pi_{1/2}, J=3/2 - 1/2$ transition}. \label{table:line-parameters}   
\centering                        
\begin{tabular}{l c c c c c c c }    
\hline\hline                 
Source                 & \multicolumn{2}{c}{Position }&  T$_{mb,RJ}$  & v$_{\rm lsr}$ & $\Delta$v & total $\tau$  & T$_{ex}$\\
                              & RA[J2000] & Dec[J2000]    &  [K]                   & \kms\             &   \kms\  &     & [K] \\   
\hline                        
   W3(OH)           & 02:27:03.90  &  61:52:24.6  &   1.83 $\pm$0.34           &  -45.70 $\pm$0.31      & 7.54 $\pm$0.87   & $0.1 - 2$ & $40.2 - 5.1$ \\         
   G10.62$-$0.39& 18:10:28.64  & -19:55:49.5  &   1.34 $\pm$0.29           &  -3.17 $\pm$0.51        &  9.50 $\pm$1.15  & $0.1 - 5$ & $30.2 - 3.7$\\ 
   NGC7538 IRS1& 23:13:45.36 &   61:28:10.5  &   1.04 $\pm$0.34           &  -57.80 $\pm$0.43      &  5.46 $\pm$1.00  & $0.1 - 5$ & $24.1 -3.5$ \\
\hline                  
\end{tabular}
\tablefoot{Line intensities are derived for the sum of the three hfs lines in each $\Lambda$-doublet of the $^2\Pi_{1/2}, J=3/2 - 1/2$ transition. We give the range of total $\tau$ values that give fits within the noise level and the corresponding T$_{ex}$ values. The total $\tau$ corresponds to the sum of optical depths of all hfs components in both doublet components. }
\end{table*}
%


\section{\label{results}Results}

\begin{figure}
\centerline{\resizebox{0.97\hsize}{!}{\includegraphics[angle=0]{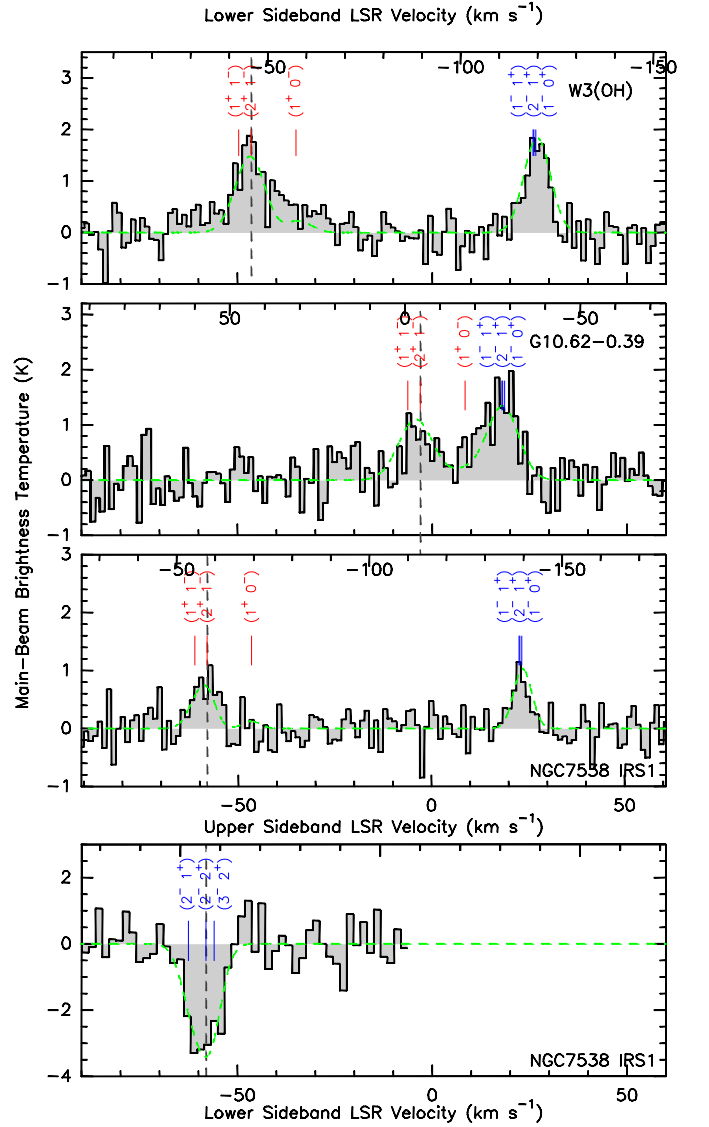}}}
\caption{Upper three panels: the $^2\Pi_{1/2}, J=3/2-1/2$ rotational lines of 
the three {\uchiir}s in emission. The gray dashed line shows v$_{lsr}$.
Red lines indicate the position of the hfs components of the - parity line of the doublet, while
blue labels show the hfs of the + transition. (Labels are shifted
for better visibility.) The dashed green line shows the $\tau=0.1$ hfs fit 
to the spectra. The lower sideband rest velocity is shifted according to the 
v$_{lsr}$ of the individual sources {and is indicated on the upper axis of each plot, while the rest velocity scale is the same for all sources}.
The lowest panel shows the $^2\Pi_{3/2}, J=3/2-5/2$ transition
toward NGC~7538~IRS1.
} 
\label{ohplot}
\end{figure}

The $\Lambda$-doublet lines of the OH $^2\Pi_{1/2}, J=3/2-1/2$ 
rotational transitions are detected, and as is characteristic of this transition, they are seen in emission (Fig.\,\ref{ohplot}) toward all three
 high-mass star-forming sites. The F = 2$^+ - 1^-$ and F = 1$^+ - 0^-$ hfs components (at 
 1837.8168 and 1837.8370 GHz, respectively)
  are separated by 3.2 \kms\  and are therefore spectrally marginally resolved. 
  The more separated and weaker F = 1$^+ - 1^-$ component (at 
 1837.7466~GHz) would be spectrally resolved,
   but lies within the noise level toward W3(OH) and  NGC~7538~IRS1. Because GREAT
    is a double-sideband receiver, the other component of the $\Lambda$-doublet 
    (a blended triplet at 1834.7355, 1834.7474, 1834.7504 GHz, for the F$=1^--1^+, 2^--1^+ , 1^--0^+$ transitions, respectively) could be recorded in the image band. For G10.62$-$0.39 the F = 1$^+ -1^-$ line partially overlaps with the emission from the image band. 

For all sources, the observed line profiles of this transition exclude the presence of a broad outflow component within our noise level of $\sim$0.4 K. The lines are turbulence-dominated and well-fitted with Gaussians within this noise level. The derived line-widths are consistent with those of other high-density tracers \citep{Plume97}, suggesting an origin from a dense, turbulent medium.


Toward NGC~7538~IRS1, the blended 2514 GHz lines of the $^2\Pi_{3/2}, J=5/2^--3/2^+$ rotational transition were also  observed and are clearly seen in absorption. The triplet of lines from the other component of the $\Lambda$-doublet at 2510~GHz was not observed.
The level of continuum emission was unstable and could not be determined reliably.

Given the OH molecules's substantial dipole moment (1.66~D), these high-frequency
rotational transitions have high critical densities $\ge10^8$~cm$^{-3}$. 
As a consequence, the LTE assumption is unlikely to apply
even in the usual high-density environment of massive protostars, and FIR radiative pumping is expected to play a role in the excitation. 
Therefore ideally one should fit the individual hfs components to derive their intensity ratio, however, 
the presented spectra do not have sufficient signal-to-noise ratio for that.
We used, hence, the HFS fitting method with CLASS and simultaneously fitted all six hfs 
components
of the doublets with fixed optical depth, $\tau$. Best fits, with a residual rms within the noise, 
are summarized in Table\,\ref{table:line-parameters}. 
The spectra's quality is moderate. For example, the broad feature underlying the 1837 GHz line toward W3(OH) is caused by an imperfect spectral baseline. 

\section{Analysis: Radiative transfer modeling}

We used RATRAN~\citep{HvdT2000} to model OH emission from protostellar envelopes, where the effect of dust heating was included by adopting models from the literature (\citealp{vdT2000}; \citealp{Mueller2002}). Only for W3(OH) we present the $^2\Pi_{1/2}, J=3/2-1/2$ line intensities calculated for a compact (few arcsecond size) source associated with OH maser emission with the large velocity gradient (LVG) model of \citet{CW91}. In addition to the far-IR pumping, this takes into account line overlap effects and the hfs structure of the lines to give a detailed prediction for the cm radio lines of OH.

\subsection{W3(OH): OH emission dominated by the {\uchii} region}



 \citet{CW91} created an LVG code\footnote{This code uses the 
collisional cross-sections from \citet{Flower89}.} to model the line properties (emission, absorption and population inversion) for the cm radio lines of OH toward W3(OH), where the 4~GHz hfs lines are seen in both emission and absorption. They show that this particular combination originates from a very narrow range of physical conditions. As a first step we reproduced their model and looked at the predictions for the far-IR lines. 

The \citet{CW91} model describes a small (few arcsecond) compact source based on interferometric observations from the literature (\citealp{Guilloteau1985, Wilson90}). They 
qualitatively 
reproduce the radio line observations using 
$T_{gas}=T_{dust}=151~$K, including the presence of both internal dust and line overlap effects. 
The model uses an abundance, $\chi_{\rm OH}=2\times10^{-7}$
and a velocity gradient of 200~km~s$^{-1}$~pc$^{-1}$ { (see also App.\ref{app:lvg})}. Line overlap is taken into account and
the velocity range {available for non-local line overlap} is fixed in this model to $\Delta$V=0.7~{\kms}.
We reproduced their model and looked into the far-IR lines (Fig.\,\ref{fig:model-lvg}, upper panel). The observations agree well with 
the predictions of the model in a density range of $n_{\rm H_2}=2-3\times10^6$~cm$^{-3}$ with 
a corresponding $T_{ex}\sim50-60$~K. We note, however, that 
the model largely overpredicts the optical depth.
Using this LVG code we also tested the importance of 
internal heating by dust, i.e. far-IR radiative pumping from the dust continuum photons. As shown in Fig.\,\ref{fig:model-lvg} (lower panel) 
we find that models without a radiation field strongly underestimate the line intensity by an 
order of magnitude.
 
 Because the known hot-core source, W3(H$_2$O), is 
only 6{\arcsec} offset from W3(OH)~\citep{Wyrowski1999}, we tested whether adding 
an envelope component could influence our model results. We used RATRAN \citep{HvdT2000}
 and adopted an envelope model from \citet{vdT2000}, with a power-law density profile n(r)=n$_0$(r/r$_0$)$^{p}$ (n$_0=5.3\times$10$^4$~cm$^{-3}$, $p=-2$) and a temperature gradient with an exponent of $-0.4$, adding up to a total luminosity of 2$\times10^4$L$_{\odot}$.
We adopted an abundance of $\chi_{OH}=0.8\times10^{-8}$,
which is observed in a similar environment by \citet{Wampfler2011}.
 These models predict very low line intensities (T$_L\sim0.3$~K), therefore we conclude that for
 W3(OH) such an envelope model cannot significantly contribute to the observed line intensities.
 This result is consistent with the observed v$_{lsr}\sim-46$~{\kms} of the emission, which 
 coincide with the {\uchii} region,W3(OH), rather than the neighboring hot-core, W3(H$_2$O). 

\subsection{NGC~7538~IRS1: OH emission dominated by an envelope component}

For this source, the 2514~GHz $^2\Pi_{3/2}, J=5/2-3/2$ 
transition was also observed. The line is seen in absorption and although the absorption feature seems to be saturated, the continuum level is uncertain. Hence, the only constraint we can
deduce is that it appears in absorption. (The similar line-width of the hfs 
fits suggests that the optical depth may be similar to that of the $^2\Pi_{1/2}, J=3/2-1/2$ transition.)

We used RATRAN to model an envelope with a power-law density and temperature profile with n$_0$=5.3$\times$10$^4$cm$^{-3}$, $p=-1.0$ and a total luminosity of 1.3$\times10^5$L$_{\odot}$ following \citet{vdT2000}.
Assuming a constant abundance of $\chi_{\rm OH}=0.8\times10^{-8}$ we reproduced the 
observed emission line intensities and the absorption feature as well (Fig.\,\ref{fig:model-ratran}). This indicates that
an envelope model can explain rotational transitions of the emitting and absorbing OH gas.
 
\subsection{G10.62$-$0.39: OH emission from an envelope and a compact source}
  
 Similarly as for the other sources, we produced models with RATRAN, where 
 the physical model of the envelope is based on models of \citet{Plume97} and
 \citet{Mueller2002}, adopting a power-law density profile with an exponent $p=-2.5$, n$_0$=1.2$\times10^9$~cm$^{-3}$ and a total luminosity of 9.2$\times10^5$L$_{\odot}$.
 These calculations produce a narrower emission feature than
 observed in our OH spectrum and underestimate the line intensity, giving
  T$_L\sim0.6$~K. Either a higher OH abundance or the addition of a compact component
  corresponding to the {\uchii} region is needed. 

\section{Conclusions}
We have detected the OH $^2\Pi_{1/2}, J=3/2 - 1/2$ rotational lines
          in emission toward three classical {\uchiir}s and the $^2\Pi_{3/2}, J=3/2-5/2$ rotational line in
          absorption toward one of them, NGC~7538~IRS1.
Our modeling shows that the far-IR radiation field plays an important role in the excitation 
          conditions. Except for NGC~7538~IRS1, an envelope model with a low OH abundance 
          ($\chi_{\rm OH}=0.8\times10^{-8}$) cannot explain the observed line intensities alone; an additional compact component
          with high density and high OH abundance is needed. We have shown  for W3(OH) that the emission is dominated by the {\uchiir} and not the nearby hot-core.
Given the particular excitation properties of the radio lines of the OH molecule, a detailed modeling
 of their properties (emission, absorption, or population inversion) and observations of more rotational transitions of OH can provide more constraints on the physical conditions.

\bibliographystyle{aa}
\bibliography{KMM-short}

\begin{acknowledgements}
We are grateful to Riccardo Cesaroni for making his radiative transfer program available to us. We also thank the referee, J. Goicoechea, for a careful reading of the manuscript. This work was partially funded by the ERC Advanced Investigator Grant GLOSTAR (247078). Based on observations made with the NASA/DLR Stratospheric Observatory for Infrared Astronomy. SOFIA Science Mission Operations are conducted jointly by the Universities Space Research Association, Inc., under NASA contract NAS2-97001, and the Deutsches SOFIA Institut under DLR contract 50 OK 0901.
\end{acknowledgements}

%

\Online

\begin{appendix} 

\section{LVG model of W3(OH)}\label{app:lvg}

\begin{figure}[h]
   \centering
   \includegraphics[angle=90,width=11cm]{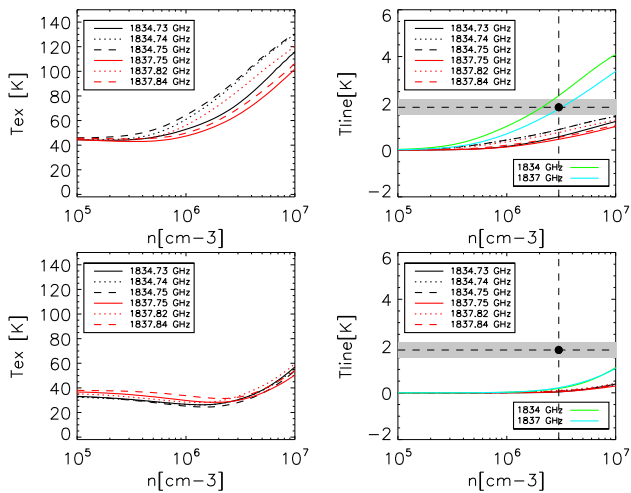}
     \caption{Results of the LVG models of \citet{CW91} showing only the predictions
     of excitation temperature and line intensity for the 1834 and 1837 GHz lines. 
      The gray dashed area
     indicates the observed line intensity including the errors. Light green and blue lines
     correspond to the sum of the hfs lines in each component of the doublet. {\bf Upper panels:}
     Models with internal radiation field. {\bf Lower panels:}
     Models without internal radiation field. }
              \label{fig:model-lvg}%
    \end{figure}

{We reproduced the model of W3(OH) of \citet{CW91}, which qualitatively models  the radio lines of OH, and show here the predictions of this model for the 1834 and 1837 GHz lines. We adopted their parameters, assuming a source size of 2{\arcsec} (0.01~pc), a velocity gradient of 200~km~s$^{-1}$~pc$^{-1}$ and an OH column density between 2$\times$10$^{14}$~cm$^{-2}$ to 2$\times$10$^{16}$~cm$^{-2}$. To compare these intensities with the observed values we used a beam size of $15{\rlap.{}''}5$ for SOFIA. In Fig.\,\ref{fig:model-lvg} two models are shown, with and without an internal radiation field introduced by the present of warm dust. As shown, the latter model underestimates the observed line intensities.}

{This code is well-adapted for the OH molecule because it includes line-overlap effects, however, for a detailed modeling the radio OH lines need to be taken into account. To perform this modeling for all the sources is beyond the scope of this letter.
}

\section{{\sl RATRAN} model of NGC7538 IRS~1}

\begin{figure}[h]
   \centering
   \includegraphics[angle=-90,width=0.65\linewidth]{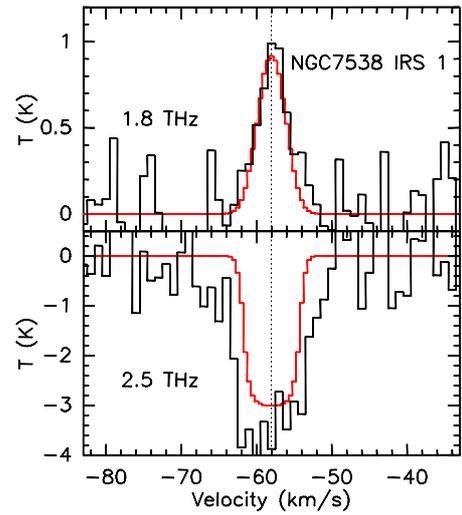}
    \caption{Our {\sl RATRAN} model of the 1834~GHz and 2514~GHz OH line in emission and in absorption, respectively. Black line shows the observed spectra with baselines removed, the red line shows the model.}
              \label{fig:model-ratran}%
    \end{figure}

{{\sl RATRAN} is based on a Monte-Carlo approach to perform radiative transfer calculations \citep{HvdT2000}. It is therefore well-suited for detailed modeling of protostellar envelopes. A radiation field from dust heating was implemented in the models. Note that as shown in Fig,\,\ref{fig:model-ratran}, a line profile can be produced. 
}

\end{appendix}

\end{document}